\documentclass{osa-article}

\journal{oe}

\articletype{Research Article}

\usepackage[version=4]{mhchem}
\usepackage{siunitx}
\usepackage{color}

\begin{document}

\title{High-Speed Phase-Only Spatial Light Modulators with Two-Dimensional Tunable Microcavity Arrays}

\author{Cheng Peng,\authormark{1} Ryan Hamerly,\authormark{1} Mohammad Soltani,\authormark{2} and Dirk R. Englund\authormark{1,*}}

\address{\authormark{1}Department of Electrical Engineering and Computer Science, Massachusetts Institute of Technology, Cambridge, MA 02139, USA\\
\authormark{2}Raytheon BBN Technologies, 10 Moulton Street, Cambridge, MA, 02138, USA}

\email{\authormark{*}englund@mit.edu} 



\begin{abstract*}
Spatial light modulators (SLMs) are central to numerous applications ranging from high-speed displays to adaptive optics, structured illumination microscopy, and holography. After decades of advances, SLM arrays based on liquid crystals can now reach large pixel counts exceeding $10^6$ with phase-only modulation with a pixel pitch of less than \SI{10}{\micro\meter} and reflectance around 75\%. However, the rather slow modulation speed in such SLMs (below hundreds of Hz) presents limitations for many applications. Here we propose an SLM architecture that can achieve high pixel count with high-resolution phase-only modulation at high speed in excess of GHz. The architecture consists of a tunable two-dimensional array of vertically oriented, one-sided microcavities that are tuned through an electro-optic material such as barium titanate (BTO). We calculate that the optimized microcavity design achieves a $\pi$ phase shift under an applied bias voltage below 10 V, while maintaining nearly constant reflection amplitude. As two model applications, we consider high-speed 2D beam steering as well as beam forming. The outlined design methodology could also benefit future design of spatial light modulators with other specifications (for example amplitude modulators). This high-speed SLM architecture promises a wide range of new applications ranging from fully tunable metasurfaces to optical computing accelerators, high-speed interconnects, true 2D phased array beam steering, and quantum computing with cold atom arrays.
\end{abstract*}

\section{Introduction}
Spatial light modulators (SLMs), which manipulate the spatial amplitude and phase distributions of light waves, find use in a wide range of applications including compact beam steering for LiDAR \cite{schwarz2010lidar,haellstig2003laser}, beam shaping \cite{forbes2016creation}, biomedical $in$-$vivo$ imaging through scattering media \cite{watts2014terahertz}, wavefront encoding for optical information processing \cite{hamerly2019large, lin2018all}, and fast programmable optical tweezers \cite{nogrette2014single, kim2019large}. Current mature commercial technologies for SLMs include liquid crystal on silicon (LCOS) and digital micromirror devices (DMDs). The slow response time of liquid crystals (LCs) limits the refresh rate to $\sim$$\SI{10}{kHz}$~\cite{zhang2014fundamentals,henderson2006free}. Micro-electromechanical systems (MEMS)-based micromirror devices are faster, with response times of tens of microseconds~\cite{dudley2003emerging,shrauger2001development}. However, the complex design and manufacturing process and the relatively high failure rate~\cite{van2003mems} due to their moving parts make them less suitable for mass production. In addition, the pixel size for both technologies is larger than \SI{100}{\micro\meter^2}.

To realize high-speed (GHz), high fill-factor spatial light modulators, several emerging SLM platforms are currently under research, including free-carrier dispersion effect modulation in silicon~\cite{soref1992silicon}, epsilon-near-zero material indium tin oxide (ITO)~\cite{huang2016gate} and quantum-confined Stark effect in semiconductors~\cite{kuo2005strong}, all of which are promising paths to high-speed amplitude modulators. However, to achieve high efficiency for applications such as high-speed display and beam steering, phase-only modulation is necessary. To date, all high-speed SLM architectures have coupled phase and amplitude modulation. A high-speed phase-only SLM design is called for.

\begin{figure}[htbp]
\centering\includegraphics[width=\textwidth]{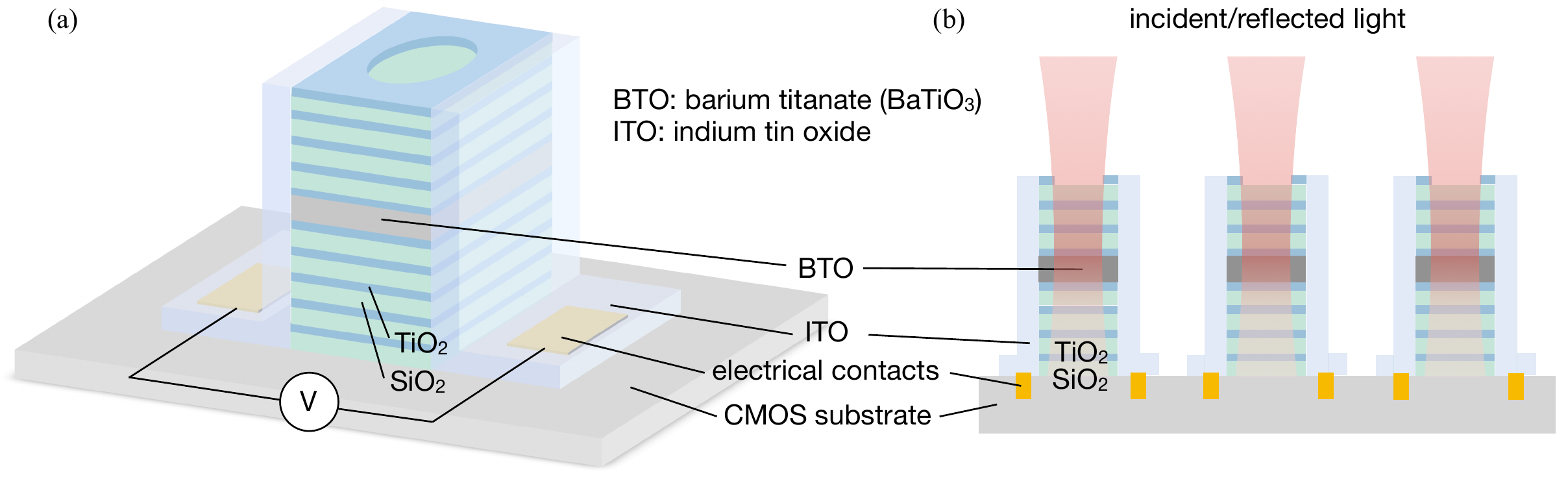}
\caption{Spatial light modulators with two-dimensional tunable microcavity arrays. (a) One phase shifter element composed of a vertical microcavity and the corresponding electrical control components. (b) An array of phase shifter elements capable of modulating the spatial profile of the reflected wavefront. \label{schematic}}
\end{figure}

In this article, we propose an SLM architecture that makes use of a two-dimensional array of vertically-oriented one-sided microcavities whose resonance frequency is modulated through an electro-optic (EO) material embedded in the cavity, as illustrated in Figure~\ref{schematic}(a). The Pockels effect shifts the refractive index of the EO material according to
\begin{equation}
    \Delta\Bigg(\frac{1}{n^2}\Bigg)_i = \Sigma_{j=1}^3r_{ij}E_j
\end{equation}
where $r_{ij}$ is the electro-optic tensor. The Pockels effect has two key properties. First, only the real part of the refractive index is modified, enabling phase-only modulation. Second, most common Pockels materials, including ferroelectric oxides, have ultrafast response times in the femtosecond range.

The downside of the Pockels effect is that it is relatively weak, so it is difficult to achieve a phase change of $\pi$ in reflection under voltages that should for practicality of driving electronics be below $\sim$$\SI{10}{V}$. To mitigate this problem, we consider the material barium titanate (\ce{BaTiO3}, also known as BTO), which has emerged as an excellent material for electro-optic modulators due to its high electro-optic coefficients experimentally characterized as large as $r_{42} = \SI{923}{\pico\meter/V}$~\cite{abel2019large}. EO modulation in BTO has sub-ps response. A Si-integrated BTO electro-optic modulator has recently been demonstrated for high speed operation up \SI{65}{GHz}~\cite{abel2019large}. The material is chemically and thermally stable and its epitaxial growth is now possible over standard silicon and silicon-on-insulator (SOI) wafers~\cite{abel2013strong,abel2019large}. But even for a high electric field of $\sim$$\SI{10}{V/\micro\meter}$, reflection through a layer thickness $h$ of BTO would yield a phase modulation change of only $\Delta\phi = \frac{2\pi}{\lambda}\Delta n(2h) \sim 0.15\pi$ for a \SI{1}{\micro\meter} film. Therefore, in the proposed SLM design, we amplify the phase change through an one-sided microcavity.

\section{Tunable microcavities as phase shifter elements}
Figure~\ref{schematic}(a) illustrates an individual vertical microcavity pixel with the EO material embedded inside the cavity. The layer of BTO has a thickness of one wavelength and is sandwiched between two distributed Bragg reflectors (DBRs) with alternating quarter-wavelength thickness layers of \ce{TiO2} and \ce{SiO2}, creating a Fabry-Perot optical cavity with strongly enhanced optical field in the BTO layer. A pair of transparent conductive oxide layers on the two opposite sides of the vertical microcavity, using materials such as indium tin oxide (ITO), form a parallel plate capacitor that generates a horizontally-oriented electric field across the BTO layer when a voltage is applied. This E-field then changes the refractive index $n$ of the electro-optic material, which in turn shifts the resonant wavelength of the optical cavity and produces a phase shift on the reflected light propagating away from the microcavity. The transparent oxide electrodes are connected to the CMOS substrate metal contacts, allowing each pixel to be individually electrically addressed with CMOS circuits.

The successful operation of the SLM with high diffraction efficiencies places two requirements on each individual phase shifter microcavity pixel: (1) the ability to have a full 0 to $2\pi$ control of the reflected light, and (2) the ability to vary the reflectance phase independently of the reflectance amplitude. To address these two requirements we design an asymmetric Fabry-Perot vertical microcavity resonator that operates in the over-coupled regime. In the section below we outline the design methodology.

\subsection{Phase shifter elements design methodology\label{cavity-design}}
The complex reflection coefficient of a one-sided resonator, calculated using the temporal coupled-mode theory (TCMT), can be expressed as
\begin{equation}\label{reflectance}
    r(\omega) = \frac{(\frac{1}{\tau_e^2}-\frac{1}{\tau_0^2})-(\omega_0-\omega)^2+2j(\omega_0-\omega)\frac{1}{\tau_e}}{(\frac{1}{\tau_e^2}+\frac{1}{\tau_0^2})+(\omega_0-\omega)^2}
\end{equation}
where $\frac{1}{\tau_0}$ and $\frac{1}{\tau_e}$ are the intrinsic loss rate of the resonator and the coupling rate between the resonator mode and the free-space mode, respectively, and $\omega_0$ is resonance frequency. This indicates that a detuning of the frequency from the resonance results in changes in the reflectance amplitude and phase. For the phase shifter resonator, as the voltage applied changes the refractive index of the active material BTO, the shifts in the resonant frequency then results in a detuning of the frequency from the resonance that modifies the amplitude and phase of the reflection wavefront as desired.

\begin{figure}[htbp]
\centering\includegraphics[width=0.85\textwidth]{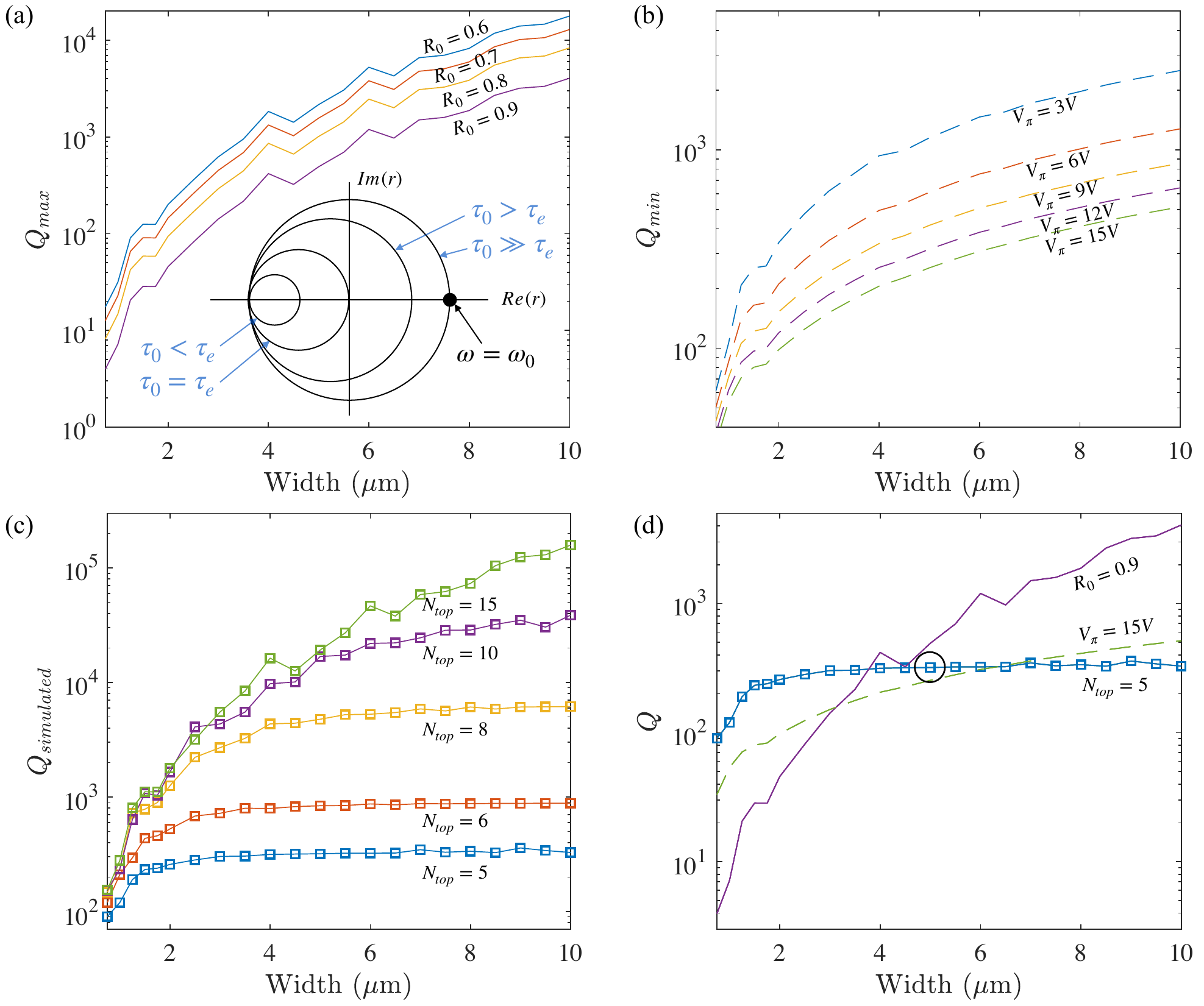}
\caption{Quality factor analysis of the microcavity resonators. (a) Maximum loaded Q for the reflectance to be greater than certain values. Inset: Schematic illustration for reflectance in different regimes of coupling between the resonator mode and the free-space mode. (b) Minimum loaded Q for $V_{\pi}$, the voltage required to achieve $\pi$ phase shift, to be less than certain values. (c) Simulation of cavity loaded Q for different micropost widths $D$ and number of \ce{TiO2}/\ce{SiO2} quarter-wavelength pairs $N_{top}$. (d) For $D = \SI{5}{\micro\meter}$ and $N_{top} = 5$, the reflectance can be maintained above $R = 0.9$ and the voltage for $\pi$ phase shift is $< 15 V$. \label{Q_analysis}}
\end{figure}

The regime of coupling between the resonator mode and the free space mode places an upper bound on the resonator's quality factor. Depending on the relative magnitude of the intrinsic loss rate $\frac{1}{\tau_0}$ and the resonator-free-space coupling rate $\frac{1}{\tau_e}$, the resonator's coupling to the free space mode can be categorized into three regimes: under-coupled ($\frac{1}{\tau_0} > \frac{1}{\tau_e}$), critically-coupled ($\frac{1}{\tau_0} = \frac{1}{\tau_e}$), and over-coupled ($\frac{1}{\tau_0} < \frac{1}{\tau_e}$) regimes. As illustrated in the inset of the Figure~\ref{Q_analysis}(a), to make sure that a full 0 to $2\pi$ reflectance phase shift and a minimal variation of amplitude can be achieved, the resonator mode should be over-coupled to the free-space mode. From Equation~(\ref{reflectance}), the reflection coefficient of the resonator at the resonance frequency $\omega = \omega_0$ can be expressed in terms of the resonator's quality factors as
\begin{equation}
    R_0 = R(\omega = \omega_0) = \Bigg(\frac{\frac{1}{\tau_e}-\frac{1}{\tau_0}}{\frac{1}{\tau_e}+\frac{1}{\tau_0}}\Bigg)^2
    = \Bigg(\frac{\frac{1}{\tau_e}+\frac{1}{\tau_0}-\frac{2}{\tau_0}}{\frac{1}{\tau_e}+\frac{1}{\tau_0}}\Bigg)^2
    = \Bigg(\frac{\frac{1}{Q_{tot}}-\frac{2}{Q_{int}}}{\frac{1}{Q_{tot}}}\Bigg)^2
\end{equation}
where $Q_{int}$ and $Q_{tot}$ are the resonator's intrinsic Q and loaded Q, respectively. For the resonator to be over-coupled to the free-space and its reflection on resonance to be greater than a given value $R_0$, this equation then places an upper bound on the resonator's loaded Q:
\begin{equation}
    Q_{tot} \leq \frac{1-\sqrt{R_0}}{2}Q_{int}
\end{equation}

On the other hand, the amount of frequency detuning that can be produced by a given applied voltage places a lower bound on the resonator's loaded Q. The frequency detuning can be expressed in terms of the change of the material's electric permittivity $\epsilon$ using perturbation theory as
\begin{equation}
    \Delta\omega = -\frac{\omega_0}{2}
    \frac{\int d^3\vec{r}\Delta\epsilon(\vec{r})|\vec{E}(\vec{r})|^2}{\int d^3\vec{r}\epsilon(\vec{r})|\vec{E}(\vec{r})|^2}
    +\mathcal{O}(\Delta\epsilon^2)
\end{equation}
The second term represents the second and higher order effects and is negligible when $|\Delta\epsilon/\epsilon|<1\%$. Since $n = \sqrt{\epsilon}$, we have $\Delta\epsilon \approx 2\epsilon\Delta n/n$. The detuning can then be written as
\begin{equation}\label{delta_omega}
\begin{split}
    \Delta\omega 
    &= -\frac{\omega_0}{2}
    \frac{\int d^3\vec{r}2\epsilon(\vec{r})\Delta n(\vec{r})/n(\vec{r})|\vec{E}(\vec{r})|^2}{\int d^3\vec{r}\epsilon(\vec{r})|\vec{E}(\vec{r})|^2}\\
    &= -\omega_0
    \frac{\int_{BTO} d^3\vec{r}\epsilon(\vec{r})\Delta n(\vec{r})/n(\vec{r})|\vec{E}(\vec{r})|^2}{\int d^3\vec{r}\epsilon(\vec{r})|\vec{E}(\vec{r})|^2}\\
    &= -\omega_0\frac{\Delta n}{n_0}
    \frac{\int_{BTO} d^3\vec{r}\epsilon(\vec{r})|\vec{E}(\vec{r})|^2}{\int d^3\vec{r}\epsilon(\vec{r})|\vec{E}(\vec{r})|^2}\\
    &= -\omega_0\frac{\Delta n}{n_0}\frac{U_{BTO}}{U_{tot}}
\end{split}
\end{equation}
where $n_0$ is the refractive of BTO without applied E field.
From Equation~(\ref{reflectance}), the detuning needed to produce a $\pi$ phase shift (from $-\pi/2$ to $\pi/2$) is
\begin{equation}
    \Delta\omega_{\pi} = 2\Bigg(\frac{1}{\tau_e}-\frac{1}{\tau_0}\Bigg)
    =\frac{\omega_0}{Q_{tot}}-\frac{2\omega_0}{Q_{int}}
\end{equation}
which according to Equation~(\ref{delta_omega}) corresponds to a refractive index change of
\begin{equation}
\begin{split}
    \Delta n_{\pi} 
    &= -n_0\frac{\Delta\omega_{\pi}}{\omega_0}\frac{U_{tot}}{U_{BTO}}\\
    &= -n_0\Bigg(\frac{1}{Q_{tot}}-\frac{2}{Q_{int}}\Bigg)\frac{U_{tot}}{U_{BTO}}
\end{split}
\end{equation}
Hence for a given $\Delta n$ to produce a phase shift greater than $\pi$,
\begin{equation}
    Q_{tot} \geq \frac{1}{\frac{\Delta n}{n_0}\frac{U_{BTO}}{U_{tot}}+\frac{2}{Q_{int}}}
\end{equation}
This is a lower bound on the resonator's loaded Q.

Figure~\ref{Q_analysis}(a) and \ref{Q_analysis}(b) illustrate  $Q_{max}$ and $Q_{min}$, the upper and lower bounds for the loaded Q, respectively, for various values of reflection on resonance and applied electrical voltages.

The intrinsic Q is calculated by simulating the resonator with perfectly reflecting DBR mirrors (15 pairs of \ce{TiO2}/\ce{SiO2} quarter-wavelength stacks) on both sides of the BTO layer. The value of the fraction of energy in the BTO layer $U_{BTO}/U_{tot}$ simulated for the intrinsic cavity is used to plot $Q_{min}$, hence representing a stricter lower bound. The $\Delta n$ corresponding to the voltage applied is calculated assuming a parallel-plate capacitor is formed by the two vertical ITO layers.

Having established the upper and lower bounds of the resonator's loaded Q, the design parameters of the resonator (the width of the vertical micropost $D$ and the number of pairs of \ce{TiO2}/\ce{SiO2} quarter-wavelength stacks in the top DBR mirror $N_{top}$) can then be selected by sweeping the parameters and searching for the optimal design that satisfies the bounding restrictions. Figure~\ref{Q_analysis}(c) shows the simulated loaded Q for resonators with various $D$ and $N_{top}$. If we narrow the design criteria to be $R_0 > 0.9$ and $V_{\pi} < 15 V$, we then end up with a set of design parameters: $D = \SI{5}{\micro\meter}$ and $N_{top} = 5$, as illustrated in Figure~\ref{Q_analysis}(d).

\subsection{Optimized phase shifter elements resonator design}
\begin{figure}[htbp]
\centering\includegraphics[width=0.85\textwidth]{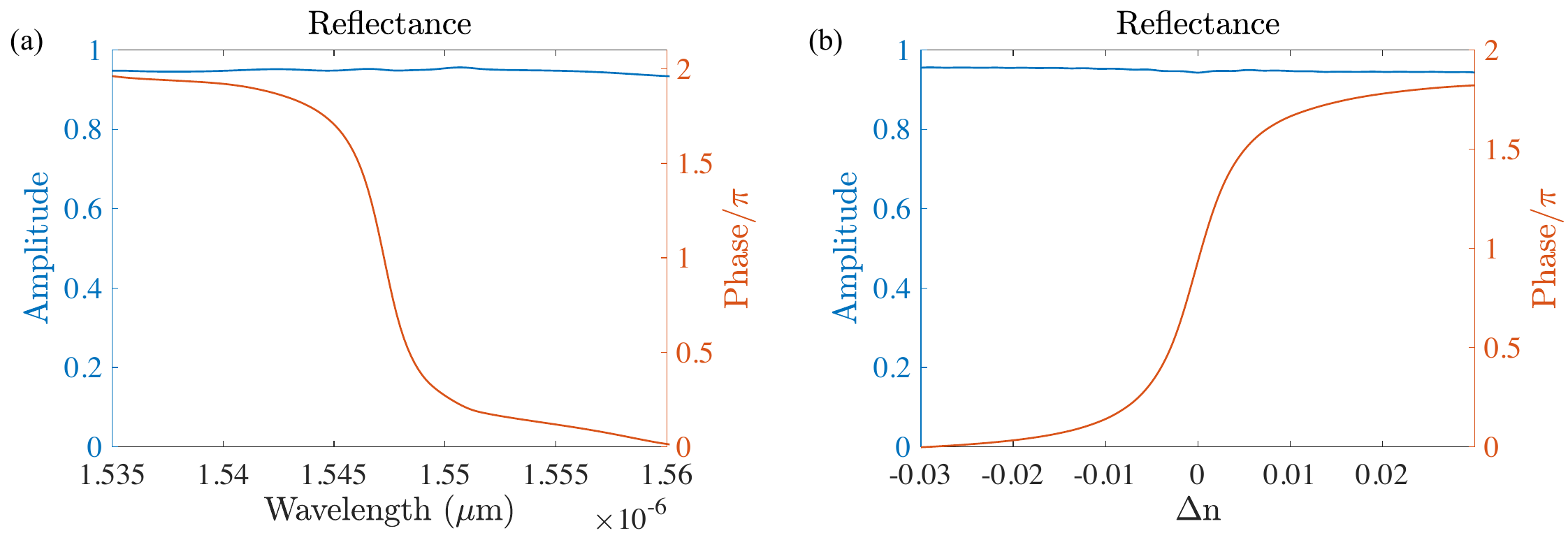}
\caption{Optimized phase shifter element with design parameters $D = \SI{5}{\micro\meter}$ and $N_{top} = 5$. (a) The reflectance spectrum. (b) The modulation characteristics as a function of the refractive index change of the active layer BTO.\label{spectrum}}
\end{figure}

The optimized phase shifter element design achieves a full 0 to $2\pi$ phase control of the reflected light while keeping the reflectance amplitude nearly constant, enabling phase-only modulation. Figure~\ref{spectrum}(a) shows the reflectance spectrum of the optimized resonator and Figure~\ref{spectrum}(b) shows the modulation characteristics of the resonator as a function of the modulated refractive index of the active material BTO.
The reflectance amplitude is maintained at $R > 0.9$ across the modulation range, and the required voltage to reach a $\pi$ phase shift (corresponding to $\Delta n \sim 0.01$) is $V_{\pi} = 9.3 V$, both agree well with the prediction from the analysis performed in Section~\ref{cavity-design}.

\begin{figure}[htbp]
\centering\includegraphics[width=\textwidth]{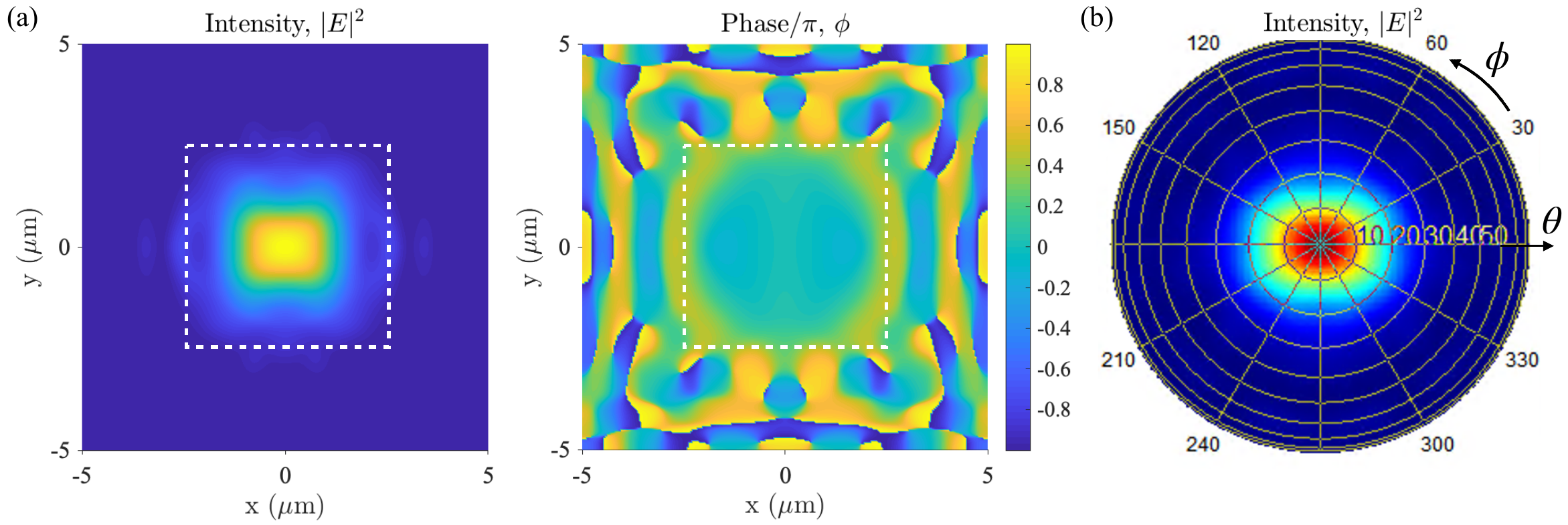}
\caption{Optimized phase shifter element with design parameters $D = \SI{5}{\micro\meter}$ and $N_{top} = 5$. (a) Near-field intensity and phase profiles of the top surface of the microcavity. The dashed box represents the outline of the microcavity's top surface. (b) Far-field intensity profile of the microcavity's radiation. \label{profile}}
\end{figure}

To facilitate coupling of the cavity mode with the free-space mode, higher-order waveguide modes that can propagate vertically in the DBR layers of the micropost should be suppressed. This can be achieved by adding an extra quarter-wavelength layer of \ce{TiO2} to the top of the micropost and etching a circular hole at the center of the layer, as illustrated in Figure~\ref{schematic}(a). This extra quarter-wavelength layer causes constructive interference at the center of the micropost and destructive interference at the peripheral areas, effectively enhancing the fundamental mode of the DBR waveguide layers and suppressing the higher-order modes. Figure~\ref{profile}(a) illustrates the E field intensity and phase profile at the top surface of the micropost.
The concentrated E field at the center of the post indicates the fundamental waveguide mode is successfully maintained as the cavity mode propagates vertically through the DBR waveguide layers. The uniform phase distribution in regions that present strong field intensities ensures that all light reflected from a single phase shifter has the same correct phase set by the applied voltage. The far-field radiation profile of this single element is shown in Figure~\ref{profile}(b), which according to Equation~\ref{farfield-equation} represents the envelope function of the far-field radiation profile of the entire two-dimensional microcavity array. The strongly concentrated field strength at the center indicates that the majority of the reflected power is distributed to the first few diffraction orders in the far field. This ensures high diffraction efficiency for most applications, where only the main diffraction lobe is of interest.

Further mode matching between the free-space mode, for example an incident light beam that has a Gaussian mode profile, and the resonator mode can be
accomplished by placing a phase mask or a microlens array in front of the two-dimensional microcavity array to change the wavefront of the incident beam to a designed wavefront profile so that the light incident on each phase shifter is matched to the waveguide mode of the input port of each microcavity.

The simulation is done with the finite-difference time-domain method using the commercially available software Lumerical FDTD Solutions~\cite{lumerical2016solutions}. The parameters used in the simulation are summarized in Table~\ref{parameter-table}. Below in Section~\ref{EO-optimization}, the determination of effective Pockels coefficient of BTO used in the simulation will be discussed in more detail.
\begin{table}
\centering
\renewcommand{\arraystretch}{1}
\begin{tabular}{ l l l l } 
Parameter & Description & Value \\ 
\hline
$n_{\ce{SiO2}}$ & refractive index of \ce{SiO2} & 1.457\cite{rii}\\
$n_{\ce{TiO2}}$ & refractive index of \ce{TiO2} & 2.3893\cite{rii} \\
$n_{\ce{BTO}}$  & refractive index of \ce{BTO}  & 2.286\cite{abel2019large} \\
$n_{b,\ce{BTO}}$ & optical birefringence of \ce{BTO}  & 0.03\cite{abel2019large} \\
$r_{42},r_{33},r_{13}$ & Pockels coefficients of \ce{BTO} & 923, 342, \SI{-63}{\pico m/V}\cite{abel2019large} \\
$d$ & thickness of the cavity layer  & \SI{678}{\nano m}  \\
$N_{\text{top}}$ & number of pairs of DBRs above the cavity layer  & 5  \\
$N_{\text{bottom}}$ & number of pairs of DBRs below the cavity layer  & 15  \\
$D$ & width of the vertical microcavity pillar  & \SI{5}{\micro m}  \\
\end{tabular}
\caption{Parameters used in simulations throughout the article.}
\label{parameter-table}
\end{table}

\subsection{Optimization of BTO's crystalline orientation\label{EO-optimization}}
The integration of BTO with the vertical microcavities, thanks to BTO's strong electro-optic (Pockels) effect, allows phase-only modulation of up to $2\pi$ under CMOS-compatible voltages. With Pockels coefficient $r_{42} = $ \SI{923}{pm/V} (in comparison, $r_{33} = $ \SI{32}{pm/V} for \ce{LiNbO3}), a large refractive index change can occur, which then translates to a substantial phase shift when only a moderate electric field is applied. Due to the crystalline structure (non-centrosymmetric tetragonal $P4mm$) of BTO, the change in the refractive index is dependent on the direction of the applied electric field. In the vertical microcavity phase shifter, the E field is applied horizontally between the two ITO contacts. In this section, we determine the optimal crystalline orientation with respect to this E field direction.

\begin{figure}[htbp]
\centering\includegraphics[width=\textwidth]{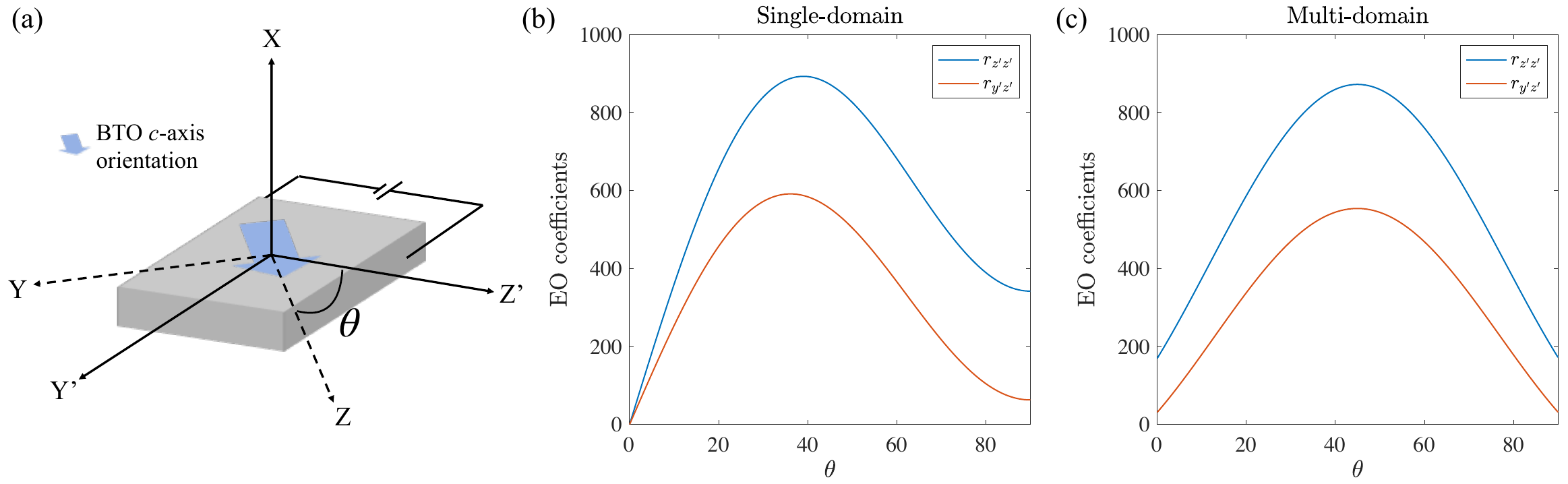}
\caption{Optimization of BTO's crystalline orientation. (a) Illustration of the BTO cavity layer, its crystalline orientation, and direction of the applied E field. (b) The electro-optic coefficients $r_{z'z'}$ and $r_{y'z'}$ as a function of the angle $\theta$ between the applied E field and the BTO $c$-axis for a single-domain BTO thin film. (c) The electro-optic coefficients $r_{z'z'}$ and $r_{y'z'}$ as a function of the angle $\theta$ between the applied E field and the BTO $c$-axis for a multi-domain BTO thin film. \label{EO_coeff}}
\end{figure}

Figure~\ref{EO_coeff}(a) illustrates the directions of the applied E field (along $Z'$-direction) and BTO's $c$-axis orientation (along $Z$-direction). We denote the angle between them by $\theta$. Since the BTO cavity layer in our resonator design has a thickness of several hundreds of nanometers, the $c$-axis of BTO is oriented in-plane~\cite{abel2013strong}. The indicatrix in the crystal's principal-axis coordinate system can be written as
\begin{equation}
    \Bigg(\frac{1}{n_o^2}+r_{13}E_z\Bigg)x^2+\Bigg(\frac{1}{n_o^2}+r_{13}E_z\Bigg)y^2+\Bigg(\frac{1}{n_e^2}+r_{33}E_z\Bigg)z^2 +(r_{42}E_y)2yz + (r_{42}E_x)2zx = 1
\end{equation}
where $n_o$ and $n_e$ are the ordinary and extraordinary refractive indices of the crystal, respectively, $r_{13}$, $r_{33}$, and $r_{42}$ are the non-zero electro-optic coefficients for $p4mm$ symmetry. The last term equals zero because the E field is applied in-plane and hence $E_x=0$. This equation can be transformed into the $X'Y'Z'$-coordinate system by substituting $x$, $y$, and $z$ by $x = x'$, $y = \cos\theta y'+\sin\theta z'$, and $z = -\sin\theta y'+\cos\theta z'$. After some algebra, the equation becomes
\begin{equation}
\begin{split}
    \Bigg(\frac{1}{n_o^2}+r_{13}E_z\Bigg)(x')^2+\Bigg[\Bigg(\frac{1}{n_o^2}+r_{13}E_z\Bigg)\sin^2\theta+\Bigg(\frac{1}{n_e^2}+r_{33}E_z\Bigg)\cos^2\theta+r_{42}E_y\cdot2\sin\theta\cos\theta\Bigg](z')^2\\
    +\Bigg[\Bigg(\frac{1}{n_o^2}+r_{13}E_z\Bigg)\cos^2\theta+\Bigg(\frac{1}{n_e^2}+r_{33}E_z\Bigg)\sin^2\theta-r_{42}E_y\cdot2\sin\theta\cos\theta\Bigg](y')^2\\
    +\Bigg[\Bigg(\frac{1}{n_o^2}+r_{13}E_z\Bigg)2\sin\theta\cos\theta-\Bigg(\frac{1}{n_e^2}+r_{33}E_z\Bigg)2\sin\theta\cos\theta+r_{42}E_y\cdot2(\cos^2\theta-\sin^2\theta)\Bigg]y'z' = 1
\end{split}
\end{equation}
Setting $E_z = 0$, we then get
\begin{equation}
    n_{z'} = \frac{n_on_e}{\sqrt{(\sin^2\theta n_e^2+\cos^2\theta n_o^2)}}
\end{equation}
which is the refractive index for light with polarization along the $Z'$-direction.
When $\vec{E} = E_z \hat{z'}$,
\begin{equation}\label{indicatrix}
\begin{split}
    \frac{1}{n_{z'}^2} \Bigg(\vec{E} = E_z \hat{z'}\Bigg)
    &= \frac{\sin^2\theta}{n_o^2}+\frac{\cos^2\theta}{n_e^2}
    + [(r_{13}+2r_{42})\cos\theta\sin^2\theta+r_{33}\cos^3\theta]E_{z'}\\
    &= \frac{1}{n_{z'}^2}\Bigg(\vec{E} = 0\Bigg)
    + [(r_{13}+2r_{42})\cos\theta\sin^2\theta+r_{33}\cos^3\theta]E_{z'}
\end{split}
\end{equation}
which then indicates that the effective electro-optic coefficient $r_{z'z'}$ for light with polarization along $Z'$-direction under E field applied in the same direction is
\begin{equation}
    r_{z'z'}(\theta) = (r_{13}+2r_{42})\cos\theta\sin^2\theta+r_{33}\cos^3\theta
\end{equation}
Similarly, the refractive index $n_{y'}$  and the electro-optic coefficient $r_{y'z'}$ for polarization perpendicular to the E field direction can be obtained using the $y'z'$ term in Equation~\ref{indicatrix}.

Figure~\ref{EO_coeff}(b) plots the effective electro-optic coefficients $r_{z'z'}$ and $r_{y'z'}$ as a function of the angle $\theta$ between the applied E field and the BTO $c$-axis, where we have used the experimentally measured values for $r_{13}$, $r_{33}$, and $r_{42}$ of BTO thin film from Ref~\cite{abel2019large}.
The maximum coefficient occurs at $\theta \approx 40^{\circ}$ for $r_{z'z'}$. Hence, to maximize the refractive index change for a given applied E field, the polarization of the light should be along the same direction as the applied E field, and the BTO crystal should be oriented such that the $c$-axis is at a $40^{\circ}$ angle from the E field direction. When multi-domain BTO thin film is considered, where the $c$-axis of the rectangular-shaped domains points in one of the two perpendicular in-plane directions, the effective $r_{z'z'}$ and $r_{y'z'}$ are the linear combinations of the EO coefficients for each single domain. Assuming the number of domains pointing in each of the two directions are equal, for symmetry reason, the effective $r_{z'z'}$ and $r_{y'z'}$ for multi-domain BTO thin film are then calculated and plotted in Figure~\ref{EO_coeff}(c). The maximum coefficient now occurs at $\theta \approx 45^{\circ}$ for $r_{z'z'}$. To maximize the refractive index change for a given applied E field, the polarization of the light should be along the same direction as the applied E field, and the BTO crystal should be oriented such that the $c$-axis is at a $45^{\circ}$ angle from the E field direction.

The maximum effective EO coefficient is $r_{z'z', max} = \SI{872.01}{\pico\meter/V}$ and the refractive index for this orientation is $n_{z'} = 2.289$. These parameters are used for simulation throughout this article.

\section{Spatial light modulators with two-dimensional tunable microcavity arrays}
The spatial light modulator, illustrated in Figure~\ref{schematic}(b), is composed of a two-dimensional array of vertical microcavities on top of a complementary metal-oxide-semiconductor (CMOS) substrate. Each vertical microcavity pixel is a free-space phase modulator that works in the reflection mode, as designed above, allowing incident light to be reflected with a phase delay which is controlled by an independent electrical voltage applied through the metal interconnect contacts in the CMOS substrate. The 2D array of vertical microcavities, when controlled independently and simultaneously, can impose a spatially variant phase distribution on the wavefront of the reflected light, which can then generate a far-field radiation pattern according to
\begin{equation}\label{farfield-equation}
\begin{split}
    U(x,y,z)
    &= \int_{-\infty}^{\infty}\int_{-\infty}^{\infty} A(f_x,f_y,0)e^{j2\pi\sqrt{\frac{1}{\lambda^2}-f_x^2-f_y^2}}e^{j2\pi(f_x x+f_y y)}df_xdf_y\\
    &= U_1(\theta,\phi,z)F_a(\theta,\phi,z)
\end{split}
\end{equation}
where $U(x,y,z)$ is the far-field radiation field of the reflected light from the SLM, $A(f_x,f_y,0)$ is the Fourier transform of the near-field radiation field of the reflected light as a function of the spatial frequencies $f_x$ and $f_y$, $U_1(\theta,\phi,z)$ is the far-field radiation field of a single pixel, $F_a(\theta,\phi,z)$ is the array factor of the system and $\lambda$ is the wavelength of light. Essentially, the far-field profile of the SLM consists of light reflected into a number of diffraction orders, the polar and azimuth angles of which are determined by the pixel pitch of the SLM. If there are many pixels in the SLM, there will be negligible overlap of different diffraction orders. The field distribution within each diffraction order can be controlled by the spatial phase distribution of the wavefront reflected from all the pixels. Using phase retrieval algorithms such as the Gerchberg-Saxton algorithm~\cite{fienup1982phase}, the required phase from each pixel can be calculated and set by the corresponding control voltage, allowing an arbitrary desired far-field pattern to be generated in each diffraction order. The power distribution of light reflected into each diffraction order is determined by the single pixel far-field pattern $U_1(\theta,\phi,z)$ which can be thought of as an ``envelope'' function.

\section{Dynamical beam steering and beam shaping}
With the optimized phase shifter elements design, we demonstrate dynamical beam steering and beam shaping using the phase-only spatial light modulator consisting of the phase shifters arranged in a large-scale two-dimensional array.

\begin{figure}[htbp]
\centering\includegraphics[width=0.75\textwidth]{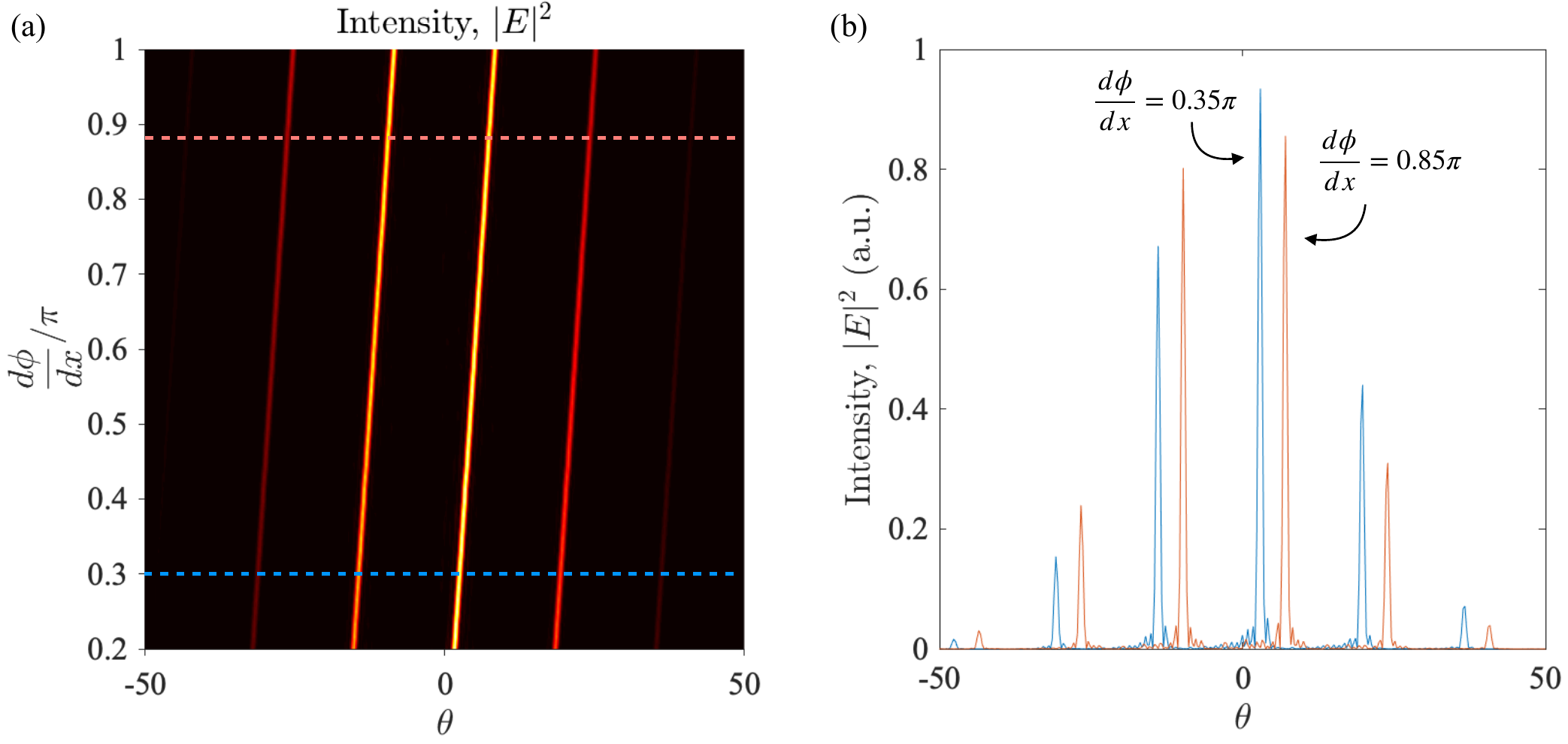}
\caption{Demonstration of dynamical continuous beam steering using a $20\times20$ array of phase shifters. (a) Far-field reflection pattern for phase profile representing phase gradients ranging from $0.2\pi$ to $\pi$. (b) Line-cuts from (a) corresponding to $\frac{d\phi}{dx} = 0.35\pi$ (blue) and $\frac{d\phi}{dx} = 0.85\pi$ (red). \label{beam-steering}}
\end{figure}

Dynamical 2D continuous beam steering can be achieved by assigning a phase profile to the shifter array that represents a phase gradient in the steering direction
\begin{equation}
    \frac{d\phi}{dx} = \frac{2\pi}{\lambda}\sin(\theta_r)
\end{equation}
where $\phi$ is the phase of the reflected light, $x$ is the spatial coordinate along the steering direction and $\theta_r$ is the reflection angle of the reflected light. Figure~\ref{beam-steering}(a) shows simulation of the continuous beam steering for a range of phase gradients $\frac{d\phi}{dx}$ from $0.2\pi$ to $\pi$ where the intensity of the E field $|E|^2$ in the far-field is plotted. Figure~\ref{beam-steering}(b) shows two line-cuts from Figure~\ref{beam-steering}(a), representing $\frac{d\phi}{dx} = 0.35\pi$ and $\frac{d\phi}{dx} = 0.85\pi$ respectively.
The peaks in the far-field intensity correspond to diffraction orders due to larger pitch size $a = \SI{5.2}{\micro\meter}$ of the phase shifter array compared to the wavelength of light $\lambda = \SI{1550}{\nano\meter}$ used. Power diffracted to the second or higher orders are negligible. Importantly, no unsteered beams ("ghost spots") remain at 0 phase gradient diffraction locations. This is a result of the phase-only modulation capability of the individual phase shifters which contributes to the high diffraction efficiency of the spatial light modulator design.

The far-field intensity profile is calculated by first sampling the simulated near-field profile of each phase shifter above the Nyquist limit, then combining the individual profiles to form a large-scale profile of the 2D array, and finally analytically propagating the near-field profile to the far field using Angular Spectrum Method. The simulated near-field profile of the individual phase shifter is sampled at 45 discrete phase levels in an index change range of $\Delta n = 0.06$. In this simulation, a $20\times20$ array of phase shifters are arranged into a two-dimensional array with pitch size $a = \SI{5.2}{\micro\meter}$. Negligible coupling between the individual phase shifters is confirmed by checking that the FDTD-simulated far-field profile agrees with the analytical far-field profile calculated with Angular Spectrum Method. 

\begin{figure}[htbp]
\centering\includegraphics[width=\textwidth]{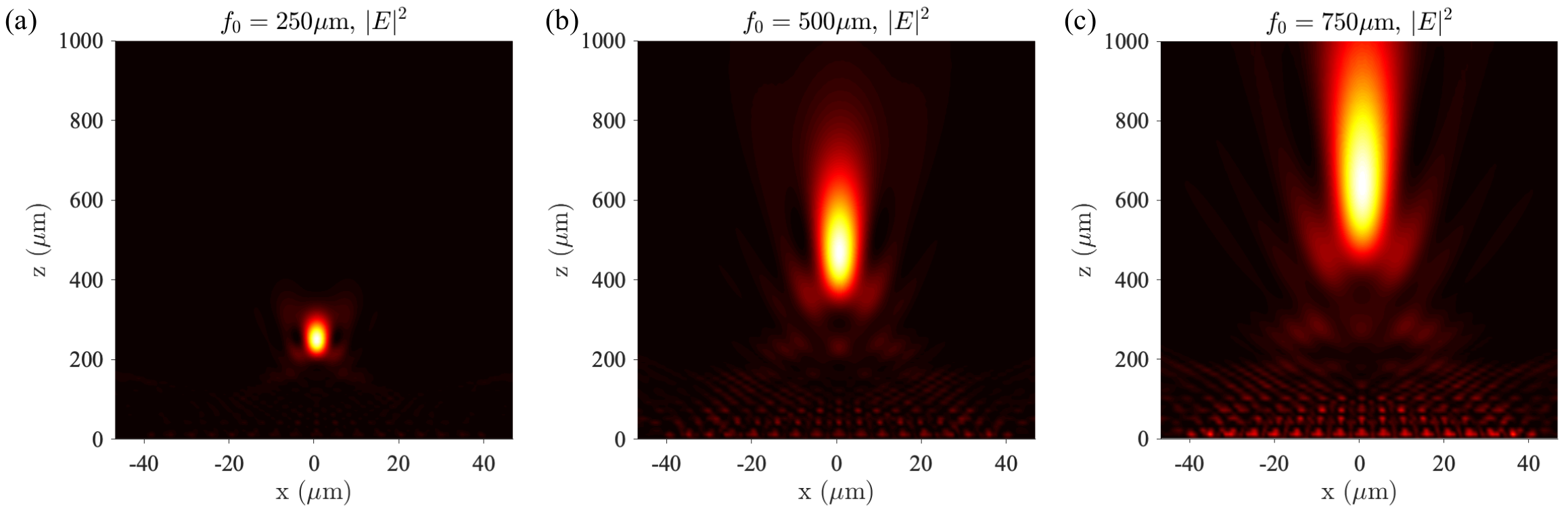}
\caption{Demonstration of dynamical beam shaping (a varifocal metalens) using a $16\times16$ array of phase shifters. Intensity profiles for focusing of light at focal lengths (a) $f_0 = \SI{250}{\micro\meter}$, (b) $f_0 = \SI{500}{\micro\meter}$, and (c) $f_0 = \SI{750}{\micro\meter}$. $z$ is the direction of reflected light propagation. \label{metalens}}
\end{figure}

Similarly, a dynamically tunable varifocal lens can be achieved by assigning a phase profile to the shifter array that follows a hyperbolic relation
\begin{equation}
    \phi(x,y) = \phi(0,0) + \frac{2\pi}{\lambda}(f_0-\sqrt{x^2+y^2+f_0^2})
\end{equation}
where $\phi(0,0)$ is the reflected phase of the center pixel and $f_0$ is the focal length of the lens. Phase profiles that correspond to different focal lengths can be dynamically assigned to the SLM pixels and the reflected light can then be focused at different distances from the SLM surface. Figure~\ref{metalens} demonstrates the dynamical tuning of the focal length of the reflected light, with the focal length set to $f_0 = \SI{250}{\micro\meter}$, $f_0 = \SI{500}{\micro\meter}$, and $f_0 = \SI{750}{\micro\meter}$, respectively.

\section{Conclusion}
In conclusion, we have proposed a high-speed, compact, phase-only spatial light modulator architecture based on two-dimensional tunable microcavity arrays, with electro-optic material BTO thin film as the active layer. The optimized microcavity design achieves simulated $\pi$ phase shift for the reflected light under an applied bias voltage of \SI{9.3}{V}, while maintaining a uniform amplitude, indicating phase-only modulation. As applications, we demonstrate voltage-tunable continuous beam deflection and a voltage-tunable varifocal lens with a two-dimensional array of the optimized microcavity resonator design. With the high diffraction efficiency enabled by the phase-only modulation, the moderate operation voltage and the high modulation speed enabled by the electro-optic effect of the ferroelectric material, as well as the experimentally feasible geometry, this SLM architecture promises a wide range of new applications ranging from fully tunable metasurfaces to optical computing accelerators~\cite{pierangeli2019large,hamerly2019large}, high-speed interconnects~\cite{kahn2017communications}, true 2D phased array beam steering, and quantum computing with cold atom arrays~\cite{bernien2017probing}.

\section*{Funding}
The research leading to these results has received funding from the US Army Research Office (Award W911NF-17-1-0435).




\bibliography{reference}

\end{document}